# A physical basis for MOND

Alasdair Macleod
University of the Highlands and Islands
Lews Castle College
Stornoway
Isle of Lewis
UK
Email: Alasdair.Macleod@lews.uhi.ac.uk

*Abstract:-* MOND is a phenomenological theory with no apparent physical justification which seems to undermine some of the basic principles that underpin established theoretical physics. It is nevertheless remarkably successful over its sphere of application and this suggests MOND may have some physical basis. It is shown here that two simple axioms pertaining to fundamental principles will reproduce the characteristic behaviour of MOND, though the axioms are in conflict with general relativistic cosmology.

## 1. Introduction

MOND is a modification to the law of gravity [1] where the acceleration is postulated to deviate from the Newtonian form at a characteristic acceleration of $a_o$ determined from observation to be of the order of $10^{-10}$ m s$^{-2}$. Specifically, if the transverse velocity with respect to the centre of mass (CM) is deduced to be $V$ from observation, the actual acceleration $a$ in a stable system is $V^2/r$ where $r$ is the distance to the CM. This is related to the expected Newtonian acceleration $a_N$ ($=\partial\Phi/\partial r$) by MOND through

$$\mu(x)a = a_N \tag{1}$$

where $x = a/a_o$ [2]. There are many effective interpolation functions, but the one commonly used for modelling the essentially flat rotation curves (RCs) of spiral galaxies is

$$\mu(x) = \frac{x}{\sqrt{1+x^2}}. \tag{2}$$

MOND scales the gravitational effect of the baryonic component at low acceleration and has been very successful at modelling the RC of galaxies with widely varying appearance and characteristics using only the mass to light ratio as a free parameter. Interpolation function (2) also anticipates the empirically valid Tully-Fisher relation.

Flat RCs are also adequately explained by a dark matter halo, and, as has already been pointed out [3], MOND is then essentially an expression for the interaction between baryonic and dark matter. However the form of the halo is parameterised to a higher degree of complexity than the simple relationship expressed by (1) and (2) – this is certainly a difficulty for the dark matter explanation. On the other hand, it has proved impossible to relate a modification to Newtonian gravity to any fundamental principle. One clue may be that $a_o$ is comparable in magnitude to $c/T$ where $T$ is the age of the Universe as derived from the Hubble parameter. If such a connection really exists, then MOND in effect incorporates into the gravitational force a time variation because $a_o$ will naturally decrease with time, and because the acceleration in the Solar System is many magnitudes larger than $a_o$, such a change is not locally detectable. But if time-varying gravitation really is a possibility, there are other ways of matching the flat RC of spiral galaxies and it is feasible to propose appropriate axioms with the same effect but which may in addition reveal fundamental principles.

In this paper we will consider how the observed RCs will emerge if it is postulated that: **1.** The expansion of the Universe is unconstrained by matter; **2.** There exists a threshold of gravitational acceleration below which an entity will join the Hubble flow. The consequences of these postulates and whether they should be considered anything other than conjecture will be briefly considered.

## 2. Time-Varying Gravitation

The proposal that the expansion of the Universe is unconstrained by the matter within, cleanly decouples the constrained and expanding frames. Consider a test object of mass $m$ within a stable circular orbit at distance $r$ from a central mass $M$ that joins the Hubble flow at absolute time $T$. We may take a simple approximation of the recession velocity $v_h$ as $r/T$ ($=H_o r$), whereupon the gravitational energy that was initially $GMm/r$ decreases after interval $\Delta T$ to $(1+\Delta T/T)^{-1}GMm/r$. Energy conservation applied to objects that participate in the expansion requires the quantity $GMm$ to increase by a balancing amount. More formally:

$$\frac{\partial(GMm)}{\partial T} = \begin{cases} 0 & \text{bound frame} \\ \dfrac{GMm}{T} & \text{expanding frame.} \end{cases} \tag{3}$$

While bearing in mind the generality of condition (3), it is reasonable to proceed on the assumption that mass remains constant through a transition into the expanding frame. Energy conservation is then guaranteed by the requirement that the relation $\partial G/\partial T = G/T$ holds between entities participating in the expansion. The expression is a first-order approximation associated with the simplified expression for the Hubble factor adopted earlier, but, for our purposes here, the approximation is adequate.

On the face of it, $\partial G/\partial T = G/T$ is a strange condition because this means the gravitational force between objects is moderated by a factor $G$ that remains constant for the duration objects remain bound, but increases with time when objects expand with respect to one another. $G$ is therefore a *relative* gravitational parameter which tracks the binding history of the entities in the same manner that spatial position difference results from the accumulative relative force between particles over all previous time. The gravitational parameter is therefore more akin to an additional dimensional variable than a constant as normally envisaged.

By this interpretation, the gravitational parameter within the Solar System is constant over time (though it may vary between constituents), being locked at the value(s) set when the component particles became gravitationally bound. In contrast, galaxies within clusters may have been unbound with respect to one another for most of the duration of the Universe and will have evolved a larger gravitational parameter. The observed dynamical behaviour of members of a galaxy cluster will then be characterised by the apparent influence of invisible matter if the lower gravitational parameter appropriate to the Solar System is incorrectly applied. Note that gravitational lensing by distant clusters will reflect the relative nature of $G$, but whilst $G$ will generally increase with distance, there is also a decrease associated with the 'look back' propagation delay.

There are also consequences on the scale of galaxies. Though the process governing the formation and development of galaxies is not completely understood, there seems little doubt that if a coalescing gas cloud possesses sufficient angular momentum, the structure will condense to a lenticular form. In the Milky Way galaxy, for example, this would comprise the nuclear bulge, halo and the thick disk component. Less certain is the origin of the so-called thin-disk component made up of much younger stars (through not exclusively so) and characterised by spiral arms. Regardless of how comes about, once the spiral galaxy is formed, subsequent galaxy mergers, it is believed, will lead to the stripping of gas and dust and the formation of relatively sterile elliptical galaxies [4]. Whilst accepting this as the most likely process by which galaxies develop, it is nevertheless possible to explore an alternative development path where the thin disk and hence the appearance of spiral galaxies is instead characterised by material *outflow* associated with the Hubble expansion. By this interpretation, the coalescence of matter into elliptical or lenticular form may be considered the first stage of galaxy formation. When the gravitational acceleration over regions in the mass distribution falls below a certain threshold, any material subject to this lower acceleration joins the Hubble flow and moves out from the centre. In a typical ellipsoid nuclear form that exhibits inner solid-body rotation, there are two points where the acceleration may simultaneously have the same minimum value; at the outer edges in the region of differential motion where the bulk of the material escapes into the plane of rotation, and also near the centre within the region of solid body motion. Because of the increase in $G$ associated with expanding entities, the outflow material will subsequently retain the same transverse centripetal velocity with the result the RC is effectively stretched in the anomalous manner that is almost universally observed. Note that near the centre, the transverse velocity matching the threshold acceleration is lower than at the outer periphery (because $a_o = V^2/r$), hence we can expect not only a dominant outflow from the outer edge that determines the properties of the RC far from the centre, but also a secondary outflow at a much lower rotational velocity originating deep within the core that perhaps traces the spiral arms[1]. Material further out from the centre has had more time to interact hence the thin disk would be expected to have a distinctive radial colour gradient, with late-type spirals becoming significantly redder with increasing radius. In addition, we expect orbits to become circular[2].

This is very much a plausibility argument based on a simple qualitative description. It fails to consider the complexity of a real dynamical process. To understand real systems, the nature of the threshold condition is key and should be explored in detail because the process, once started, must obviously become more-or-less self-sustaining because spiral galaxies happen to be very common in the Universe. But the most plausible threshold condition does not seem very promising in terms of sustainability: If the threshold acceleration is an inverse function of time such as $c/T$, it is uncertain how material that is above the threshold at any time can later get beneath the falling threshold. But whatever the mechanism, one might expect blips, perhaps related to discontinuities in the core mass distribution, where the outflow will cease. This would show up as rings in otherwise normal spiral galaxies as the outflow toggles on and off, thus a detailed analysis of galaxies with this type of morphology or features might give a strong insight into the threshold condition. Indeed many such galaxies do exist; a good example is Hoag's galaxy which displays a dramatic transition [7]. A review by Buta and Combes [8] found that as many as a fifth of all spiral galaxies display a ring or multiple ring structure, and though collisions and resonance effects associated with bars can explain many, there are some examples for which a convincing explanation is not forthcoming.

Not all galaxies are spirals. Elliptical galaxies do not exhibit solid-body rotation and as the bulge becomes more spherical, material outflow is no longer confined to a single thin plane and can reduce the density of star-forming material to actually inhibit star formation. This type of distribution would not produce an observable spiral or disk. In fact the elliptical classification E7 may be the limit above which the escaping material is of such low density it becomes unable to form stars.

## 3. Derivation of Rotation Curves

In spite of reservations about the nature of the threshold, it is possible to illustrate how the observed range of surface density profiles and the anomalous forms of RCs can be reproduced by constructing a basic model The radial surface density associated with an elliptical galaxy, or the nuclear bulge of a spiral galaxy, is best described by the empirical de Vaucouleur's Law, but for numerical modelling the simpler form of the Hubble function is preferable. A problem with the Hubble expression is that it does not incorporate the solid-body motion often present at the core of spirals and barred spirals. Consequently, the expression is modified here to give the possibility of solid-body rotation at small radius and the conventional Hubble form at large radius:

$$\rho(r) = \rho(0) \frac{r/a}{(1+r/a)^3}. \qquad (4)$$

A central point source (a black hole) can generally be included, but is set to zero in this simulation. By an appropriate selection of a central surface density $\rho(0)$, characteristic length $a$, and cut-off radial extent, it is possible to generate a range of arbitrary distributions. If we consider a galaxy of radius 10 kpc and normalised with total mass of $10^{11}$ $M_\odot$, there are a number of distributions possible. Fig. 1 shows the rotational velocity and it is clear that rising or falling curves, with or without a peak, are typical. The characteristic acceleration under the assumption of spherical symmetry is also shown.

---

[1] This is evident from the Whirlpool galaxy for example, where the spiral arms are seen to originate from deep within the core [5]. Bear in mind though that the outflow velocity is reduced by the closeness to the CM.

[2] It has been suggested that entities with proper motion with respect to the Hubble flow (i.e. in the radial direction) will be subject to a retardation that will dissipate the motion [6].

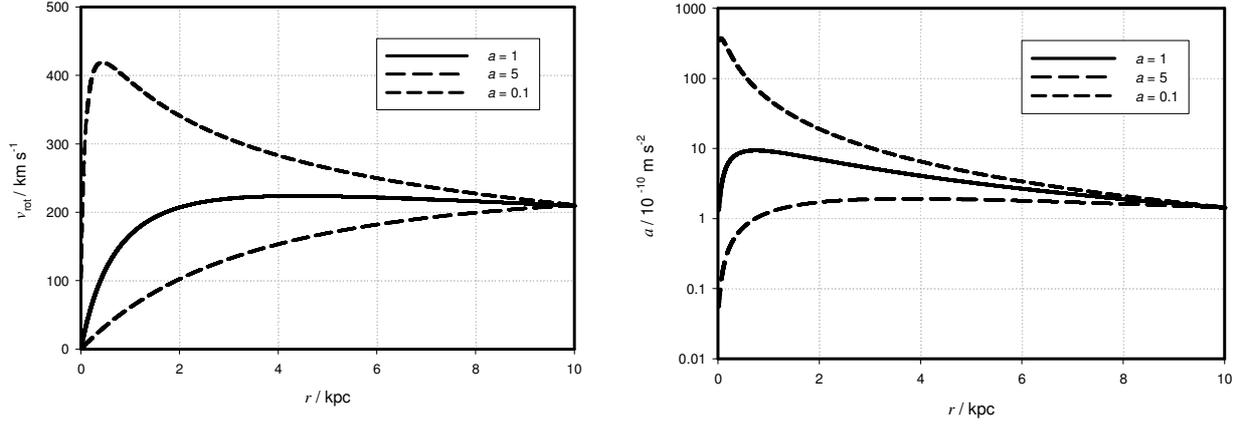

**Figure 1** A nuclear bulge of radius 10 kpc and total mass $10^{11}$ $M_\odot$ plotted with various values of $a$ in equation (1). The left hand graph shows the rotational velocity arising from the mass distribution and the right hand graph shows the corresponding acceleration (in units of $10^{-10}$ m s$^{-2}$).

We can construct a model on the assumption that at a particular time ($T = 6$ byr in this particular example to reflect the probable age of the disk of the Milky Way galaxy) the threshold limitation suddenly becomes active and develops like $c/T$. The shape of the RCs over time can be predicted for each of the distributions of Fig. 1. Near mass influences that give rise to the small-scale variations on the RCs are ignored, but of course the time development of $G$ is included. The results are shown in Fig. 2.

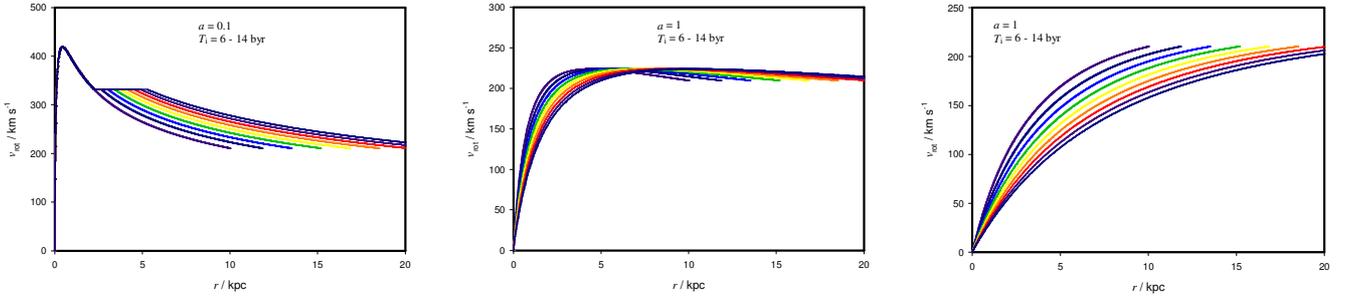

**Figure 2** The development of the rotation curve in 1 byr steps up to 14 byr when the threshold condition is applied when the Universe is 6 byr old. The different values of $a$ permit falling, flat and rising curves outwith the core ($r > 10$ kpc) in a region where there is very little material and Keplerian orbits are expected from a straightforward application of Newton's laws. (The curves proceed from left to right over time.)

The extended RC is seen to be the stretching out of the core RC that existed at a much earlier time with, in some cases, the additional complication of an outflow from deep within the core. In spite of our earlier reservation about the sustainability of the effectiveness of a threshold decreasing with time, it is immediately clear the process can potentially be self-sustaining because material lost from the centre will lead to a general reduction in acceleration throughout the galaxy, possibly at a greater rate than the threshold acceleration falls with time (though the models above are too crude to be much influenced by adopting a constant threshold level or even a level *increasing* with time). The flat or increasing slope that is frequently observed then merely reflects the original mass distribution within the bulge (whose shape is recovered by running the process backwards in time and exploiting the fact that the transverse *and* radial velocities of the escaping matter remain constant).

It would be useful to match the revolution curves of specific galaxies to the model, but it is unnecessary to do so – we merely have to show the equations associated with MOND arise from the two postulates and are completely consistent with them in the domain of spiral galaxies. The effectiveness of MOND in these situations has already been demonstrated. When $a_N \ll a$, as is the case at the outer reaches of the disk, from (1) and (2) the MOND acceleration tends to

$$a = \sqrt{a_o\, a_N} \quad \propto r^{-1} \tag{5}$$

In the case presented here, the first postulate states that an entity participating in the expansion maintains a constant potential over time (this is just energy conservation), thus

$$\frac{GM}{r} = constant \quad \Rightarrow a \propto r^{-1}, \tag{6}$$

where $M$ is the mass of the galaxy internal to the orbit, remembering $G$ is varying like $r$. The MOND modified acceleration and the first postulate are therefore consistent. MOND makes a further effective prediction: With MOND, the observed rotational velocity, from (5) is

$$V^4 = r^2 a_o a_N = (Ga_o)M, \tag{7}$$

the Tully-Fisher mass relation. This connects the rotational velocity at the full extent of the disk to total mass. In our alternative representation, the rotational velocity at the full extent of the disk was acquired when material at the outer edge of the disk escaped the gravitational force of a bulge (which included all the interior mass of the disk). The local Newtonian acceleration $a_h$ at that time matched the threshold acceleration at which mass joins the Hubble flow. Thus the second postulate implies

$$V^4 = r^2 a_h^2 = (Ga_h)M. \qquad (8)$$

If we identify $a_o$ in MOND with the threshold acceleration on which a mass joins the Hubble flow, the expressions become identical.

We can therefore identify the characteristic equation of MOND, (1), as a single mathematical expression encompassing the principle that gravitationally bound masses join the Hubble flow once the acceleration falls to $a_o$ and on doing so, the gravitational parameter $G$ subsequently increments with time by $G\partial t/T$.

Moffat's non-symmetric gravity theory also attempts to explain the flat RCs [9] by proposing an increase in gravitational force associated with a modification to general relativity, but the parameter choice again has no clear basis. The characteristic equation is

$$G(r) = G_o \left( 1 + \sqrt{\frac{M_o}{M}} \left[ 1 - \left(1 + \frac{r}{r_o}\right) e^{-r/r_o} \right] \right), \qquad (9)$$

where the subscript refers to constant values. Differentiating twice and lumping the constants together as $A$,

$$\frac{\partial^2 G(r)}{\partial r^2} = A \left(1 - \frac{r}{r_o}\right) e^{-r/r_o}, \qquad (10)$$

the flat region of the curve is selected by setting $r_o$ to that value, typically 13.5 kpc. But it is not as effective as MOND.

## 4. Discussion

It has been proposed in this paper that MOND is really a description of the process by which a body joins the Hubble flow and the subsequent stipulation of energy conservation. With this modification MOND no longer implies a change in inertia, a change to Newton's second law, or even an adjustment to the Newtonian theory of gravity. Special relativity and general relativity are unaffected because what is being described is merely a boundary transition effect. If one is looking at a physical basis for MOND, this is certainly a possibility, though not perhaps to everyone's taste: There may be big pluses, but the cost is high. All of general relativistic cosmology is lost. The accepted cosmological model is based on an assumption of coupling between expansion and matter: Matter slows the expansion because kinetic energy is depleted by supplying the energy needed to expand against the gravitational field. This is the precise opposite to the first postulate which states there is complete decoupling. Energy conservation is ensured by the increase in the gravitational parameter which keeps the gravitational potential constant. Thus the Universe expands – we do not know why or precisely at what rate, but what is certain is that the rate is unaffected by the matter content of the Universe.

Does basing MOND on these new principles have any basis in reality, or is it even useful? Particularly as it seems to raise a very difficult new question – why should there be an acceleration limit associated with a transition to the expanding frame? This problem has already been explored inconclusively [6], where it was suggested an energy limit would make some sense. Nevertheless the separation of the bound and expanding frames is intriguing, and it would be useful to obtain detailed information about the transition acceleration as it varies with time, application domain and possibly even mass. This data can be obtained through a detailed analysis of the properties of ring (and ringed) galaxies. Information is also available from the way $a_o$ is seen to change in different circumstances within the MOND description (relaxing the condition that $a_o$ must be a constant).

## References

[1]   Milgrom, M. 1983, "A modification of the Newtonian dynamics as a possible alternative to the hidden mass hypothesis", ApJ, **270**, 365
[2]   McGaugh, S.S. 2004, "The mass discrepancy-acceleration relation: Disk mass and the dark matter distribution", ApJ, **609**, 65
[3]   Famaey, B., Gentile, G., Bruneton, J.-P. & Zhao H.S. 2007, "Insight into the baryon-gravity relation in galaxies", Phys. Rev. D75, 063002
[4]   Seeds M.A. 1999, *Foundations of Astronomy*, Ch. 17, WADSWORTH
[5]   *Ibid.*, Fig. 16-25, p 325
[6]   Macleod. A. 2006, "Physics at the transition between bounded and unbounded trajectories", arXiv:physics/0611091
[7]   http://antwrp.gsfc.nasa.gov/apod/ap020909.html, Accessed 8th August 2007
[8]   Buta, R. & Combes. F. 1996, "Galactic rings", Fund. Cosmic. Phys., **17**, 95.
[9]   Moffat J.W. 2004, "Gravitational theory, galaxy rotation curves and cosmology without dark matter", arXiv:astro-ph/0412195